\documentclass[%
 reprint,
superscriptaddress,
 amsmath,amssymb,
 aps,
prb,
]{revtex4-2}

\usepackage{graphicx}
\usepackage{dcolumn}
\usepackage{bm}
\usepackage{tabularx}
\usepackage{braket}
\usepackage{xtab,afterpage}
\usepackage{multirow}
\usepackage{booktabs}
\usepackage{xcolor}
\usepackage[normalem]{ulem} 
\usepackage{comment}

\def\ypo4{YPO$_4$}
\def\d#1/d#2{\frac{\partial #1}{\partial #2}}

\def\basisu#1s#2p#3d{\hbox{\it (#1s#2p#3d)\ }}
\def\basis#1s#2p#3d/#4s#5p#6d{\hbox{\it (#1s#2p#3d)/[#4s#5p#6d]\ }}

\usepackage{soul}

\def\ypo4{YPO$_4$}
\newcounter{tablfootnote}
\def\fn#1{\setcounter{tablfootnote}{#1}%
          \raise0.5ex\hbox{\ \footnotesize{\alph{tablfootnote}%
}}}

\makeatletter
\def\yo8{YO$_8$@CTEP$_{\rm min}$}
\def\ypo46{Y(PO$_4$)$_6$@CTEP$_{\rm ext}$}
\makeatother
\newcounter{enumsli}
\def\sli#1{\setcounter{enumsli}{#1}(\roman{enumsli})~}

\begin{document}

\title{Compound-tunable embedding potential method to model local electronic excitations on $f$-element ions in solids: Pilot relativistic coupled cluster study of Ce and Th impurities in yttrium orthophosphate, YPO$_4$}

\author{A.~V.~Oleynichenko}
\email{oleynichenko\_av@pnpi.nrcki.ru}

\affiliation{B. P. Konstantinov Petersburg Nuclear Physics Institute of National Research Center ``Kurchatov Institute'' (NRC ``Kurchatov Institute'' -- PNPI), Gatchina, Leningrad district 188300, Russia}

\affiliation{Moscow Institute of Physics and Technologies (National Research University), Institutskij pereulok 9, Dolgoprudny Moscow region 141700, Russia}

\author{Y.~V.~Lomachuk}
\email{lomachuk\_yv@pnpi.nrcki.ru}

\author{D.~A.~Maltsev}
\author{N.~S.~Mosyagin}
\author{V.~M.~Shakhova}

\affiliation{B. P. Konstantinov Petersburg Nuclear Physics Institute of National Research Center ``Kurchatov Institute'' (NRC ``Kurchatov Institute'' -- PNPI), Gatchina, Leningrad district 188300, Russia}

\author{A.~Zaitsevskii}

\affiliation{B. P. Konstantinov Petersburg Nuclear Physics Institute of National Research Center ``Kurchatov Institute'' (NRC ``Kurchatov Institute'' -- PNPI), Gatchina, Leningrad district 188300, Russia}

\affiliation{Department of Chemistry, M. V. Lomonosov Moscow State University, Leninskie Gory 1/3, Moscow 119991, Russia}

\author{A.~V.~Titov}
\email{titov\_av@pnpi.nrcki.ru}

\affiliation{B. P. Konstantinov Petersburg Nuclear Physics Institute of National Research Center ``Kurchatov Institute'' (NRC ``Kurchatov Institute'' -- PNPI), Gatchina, Leningrad district 188300, Russia}

\affiliation{Saint Petersburg State University, 7/9 Universitetskaya nab., 199034 St. Petersburg, Russia}

\date{\today}

\begin{abstract}
A method to simulate local properties and processes in crystals with impurities
via constructing cluster models within the frame of the compound-tunable embedding potential (CTEP) and highly-accurate {\it ab initio} relativistic molecular-type electronic structure calculations is developed and applied to the Ce and Th-doped yttrium orthophosphate crystals, YPO$_4$, having xenotime structure.
Two embedded cluster models are considered, the ``minimal'' one, YO$_8$@CTEP$_{\rm min}$, consisting of the central Y$^{3+}$ cation and its first coordination sphere of eight O$^{2-}$ anions (i.~e.\ with broken P--O bonds), and its extended counterpart, Y(PO$_4$)$_6$@CTEP$_{\rm ext}$, implying the full treatment of all atoms of the PO$_4^{3-}$ anions nearest to the central Y$^{3+}$ cation. CTEP$_{\rm min,ext}$ denote here the corresponding cluster environment described within the CTEP method.
The relativistic Fock-space coupled cluster (FS RCC) theory is applied to the minimal cluster model to study electronic excitations localized on Ce$^{3+}$ and Th$^{3+}$ impurity ions. Calculated transition energies for the cerium-doped xenotime are in a good agreement with the available experimental data (mean absolute deviation of ca.~0.3~eV for $4f{\to}5d$ type transitions).
For the thorium-doped crystal the picture of electronic states is predicted to be quite complicated, the ground state is expected to be of the $6d$ character. The uncertainty for the excitation energies of thorium-doped xenotime is estimated to be within 0.35~eV.
Radiative lifetimes of excited states are calculated at the FS RCC level for both doped crystals. The calculated lifetime of the lowest $5d$ state of Ce$^{3+}$ differs from the experimentally measured one by no more than twice.
\end{abstract}

\maketitle

\section{Introduction}

{\it Ab initio} simulation of electronic structure and properties of crystals based on periodic models is a well-developed branch of solid state physics (see \cite{Zhang:19, McClain:17, Evarestov:07} and references therein). These models work fine for crystals of pure compounds, but meet difficulties in modeling local impurities and defects as well as processes like electronic excitations localized on a single impurity atom or processes of energy or charge transfer between two or more impurity centers. \textit{Ab initio} periodic structure models with extended unit cells of reasonable size cannot provide the accuracy usually required for such systems. More importantly, one can argue that rigorous many-body methods like coupled cluster (CC) theory formulated for periodic-structure systems are still not feasible for crystals with complicated structure and containing heavy $d$- or $f$-element atoms.

A natural way towards highly reliable quantum-mechanical studies of electronic states and properties of impurity centers in crystals consists in constructing cluster models~\cite{Kantorovich:88,Grimes:92}. Such models allow one to apply the most accurate methods of modern electronic structure theory as well as high-performance computer codes, accurate small-core pseudopotentials (PPs) and flexible basis sets to study crystal fragment properties in the presence of impurity atoms with high accuracy. Recent reviews of modern approaches to constructing cluster models can be found in~\cite{Ghosh:18,Maltsev:21_prb}. Embedding potential based methods are among the most well-established approaches to account for the effects of a crystal environment in a molecular-type cluster study; they also seem to be the most successful in simulating local electronic excitation processes. For example, the frozen density embedding method developed in the series of papers~\cite{Huang:06,Gomes:08,Hofener:12, Gomes:12,Hofener:13} (see also references therein), being an approach of the WFT-in-DFT type (wavefunction-based method in density functional theory) was applied to study solvatochromic shifts in light molecules~\cite{Hofener:13} and coupled excitations in several organic molecules~\cite{Hofener:16} by the CC response theory. The notable application of the WFT-in-DFT methodology directly connected to the subject of the present work was presented in~\cite{Gomes:08}, where it was successfully used in conjunction with the FS RCC theory to calculate excitation energies of the $5f$-$5f$ transitions located at the Np$^{6+}$ impurity ion embedded in a Cs$_2$UO$_2$Cl$_4$ crystal. However, nearly all embedding potentials proposed to date do not allow one to exploit the cluster models of sizes amenable for high-level WFT modeling, especially if covalent-type bonds are to be broken. Moreover, the approaches cited above are rather theoretically involved and often require the use of specific software.

To bypass the complicating factors mentioned above and pave the way towards highly accurate cluster modeling of local processes and impurity sites in solids, the concept of the compound-tunable embedding potential (CTEP) was proposed recently in the series of papers~\cite{Lomachuk:20_pccp,Maltsev:21_prb, Shakhova:22}. Within this approach the impurity atom and its first coordination sphere (the so-called main cluster) are treated at the best possible level of theory, while atoms from the layers next to the main cluster are simulated using shape-consistent PP operators of a special type and fractional point charges. In principle, more than one impurity site can be considered~\cite{Maltsev:21_prb} to simulate rather complicated and delocalized processes. The CTEP methodology can be regarded as a logical continuation of the ideas underlying the pseudopotential theory (see \cite{Titov:99, Mosyagin:06amin} and references therein). For this reason it is capable to simulate (approximately but quite efficiently) the exchange interaction between the main cluster and its environment when approaches of different types like WFT and DFT are used to describe these two parts of the system~\cite{Maltsev:21_prb}, unlike other embedding theories.

Some basic justification for such a choice is as follows. Long-range Coulomb potentials from atoms of environment acting on electrons of the main cluster (where these potentials are well-behaved harmonic functions) are naturally represented by large-core pseudopotentials of the nearest environmental cations (NCE, see below) augmented by a system of fractional point charges (simulating mainly the contributions from the Coulomb field of distant atoms) since they are described by a well-known local operator. In contrast, the exchange part of an embedding potential should be in principle described by an nonlocal operator and the ability of semilocal shape-consistent PPs to reliably simulate this part is not a priori obvious. However, due to optimization of the embedding pseudopotential parameters using calculations of a crystal (rather than atomic or fragment ones) the resulting semilocal representation of the nonlocal exchange operator is expected to be rather accurate at least for describing the ground state of a considered pure crystal. This is illustrated, for instance, by the comparison of electron densities obtained with the cluster CTEP model of YPO$_4$ and within periodic structure calculation (Fig.~\ref{fig:densities}).
Taking into account the short-range nature of the exchange operator and a good transferability of shape-consistent PPs (see \cite{Goedecker:92a} and section ``Theory'' in \cite{Titov:99}), our CTEP works well also for crystals with impurities \cite{Lomachuk:20_pccp, Maltsev:21_prb} and for excited states in the cases where the excitations are well localized on the central atom, i.~e.\ quite far from the main cluster boundary (see Table~\ref{tab:exc-fscc}).

We consider the CTEP approach as an alternative to the well-known embedding \emph{ab initio} model potential (AIMP) methods, both in general formulation (see \cite{Barandiaran:88_AIMP-NaCl:Cu+, Francisco:92_Huzinaga-Adams, Ruiperez:05_Cs2NaYCl6:Ce-Tb, Seijo:04_AIMP-review} and references therein) and in its fragment (FAIMP) version  (see \cite{Swerts:08_FAIMP, Larsson:23_FAIMP+Ce:YVO4} and references), analogously to the semilocal norm-conserving/shape-consistent PPs \cite{Ermler:88, Stoll:02} and Gatchina (generalized) relativistic PPs (GRPPs) \cite{Titov:99, Mosyagin:06amin, Oleynichenko:2023} vs.\ AIMP core potentials \cite{Huzinaga:71, Huzinaga:87_AIMP, Seijo:04_AIMP-review}.

The CTEP method was shown to work well for both ionic crystals like halogenides and niobates~\cite{Maltsev:21_prb} and crystals including complex anions with covalent bonds like orthophosphates of $d$-element (yttrium) with $f$-element ions as impurities (Th and U)~\cite{Lomachuk:20_pccp}. However, the cluster models constructed in the mentioned papers were still too large for accurate {\it ab initio} treatment of correlations and effects of special relativity, so the development of even more compact models is desirable. For example, the cluster model of xenotime YPO$_4$ presented in~\cite{Lomachuk:20_pccp} involved several dozens of explicitly treated atoms with $\sim$300 electrons. To bypass this difficulty and make it possible to use CC-based electronic structure methods, in the present work the concept of a ``minimal cluster with broken covalent bonds'' in complex anions (P--O bonds of the orthophosphate ions PO$_4^{3-}$ in our case) is presented, that allows one to dramatically reduce the number of explicitly treated atoms (and, hence, electrons). 

Since the electronic shells in the original YPO$_4$ crystal are closed, to reproduce electronic and spin densities in the vicinity of the central Y atom (Y$_c$), all the oxygen anions within our minimal CTEP model are considered as having completely occupied shells (1s$^2$2s$^2$2p$^6$). However, their charge distributions are distorted, due to both the ionic Y$_c^{3+}$–O$^{2-}$ bonding with explicitly treated Y$_c^{3+}$ cation and covalent bonding with the nearest six P atoms. Distortions also arise from the ionic bonding with more distant four Y cations described by CTPPs and from the effect of partial charges on the environmental pseudoatoms. The correct description of electronic densities on P and Y cations outside of the minimal cluster is not important for this model, since these CTPPs and partial charges are generated to reproduce only the electronic densities within the main cluster as it demonstrated in Fig.~\ref{fig:densities} (an analogy with the norm-conserving PP smoothing the orbital in the core region while reproducing the correct behavior in the valence region would be appropriate). Removing/adding some electrons from/to the given minimal CTEP model would result in appearance of the nonzero spin density in the main cluster that would be in contradiction with the periodic structure studies and experiments.
As a consequence, the computational cost of accurate wavefunction based methods can now be acceptable for such a minimal CTEP model with broken covalent bonds to study properties not attainable within the DFT based studies.

The present paper reports pilot applications of the CTEP approach with broken covalent bonds combined with the relativistic coupled cluster (RCC) correlation treatment to the modeling of electronic excitations localized on the $f$-element impurity ions Ce$^{3+}$ and Th$^{3+}$ in solids. For the pilot study the xenotime (yttrium orthophosphate, YPO$_4$) crystalline matrix was chosen. Due to the high symmetry of the crystal structure of this compound (tetragonal, $I4_1/amd$; Y$^{3+}$ ions occupy the sites with the $D_{2d}$ local symmetry), quantum-chemical calculations of its properties are expected to be less computationally demanding compared to other orthophosphates, in particular, with monazite-like structures.

Xenotime-type crystals doped with different lanthanide ions possess unique optical properties, the scope of practical applications includes new efficient tunable solid-state lasers, luminescent materials, scintillators, phosphors and amplifiers for fiber-optic communication (see~\cite{Laroche:01,Lai:07,Gupta:21} and references therein). From the theoretical point of view, the Ce$^{3+}$-doped xenotime is the simplest material of this type, since its spectrum is determined mainly by the single electron occupying either $4f$ or $5d$ orbitals of the cerium ion. Optical properties of YPO$_4$:Ce$^{3+}$ were extensively studied in the last few decades, and the experimental data on excited states and transition energies due to the cerium impurity are available~\cite{Karanjikar:88,Sytsma:93,Laroche:01,Pieterson:02,Lai:07,Luo:09,Krumpel:09,Zhan:12, Dorenbos:13,Kahouadji:19}, allowing one to assess the accuracy of the developed CTEP+RCC approach.

Xenotime is a material with high chemical and radiation resistance and is considered (\cite{Urusov:12,Cutts:16} and references therein), among other orthophosphates, as a natural analog of matrices for a long-term actinide immobilization. These remarkable features along with a very large band gap of perfect xenotime ($>$\,8.6~eV\,\cite{Makhov:02,Wang:09} exceeding the $^{229m}$Th nuclear excitation energy ($\sim$8.3~eV~\cite{Kraemer:23}) allows one to consider xenotime as a very prospective matrix for a solid-state nuclear clock~\cite{Kozlov:23}. To the authors' best knowledge neither experimental nor theoretical data on the thorium-doped xenotime YPO$_4$:Th$^{3+}$ were previously published, except for our previous paper on the CTEP application~\cite{Lomachuk:20_pccp}, where the thorium impurity oxidation state, geometry and vibrational modes of the defect site were studied at the DFT level.

The paper is organized as follows. First we briefly outline the main features of the CTEP concept and present the method of constructing the ``minimal'' cluster models. Then the relativistic Fock space coupled cluster theory used to calculate excitation energies and radiative lifetimes of excited states is reviewed. Afterwards, the minimal cluster model of xenotime is constructed and the accuracy of the model is assessed. Then the results of relativistic coupled cluster calculations of localized low-energy electronic excitations in the YPO$_4$:Ce$^{3+}$ and YPO$_4$:Th$^{3+}$ doped crystals are presented, the accuracy and the scope of applicability of the novel CTEP+RCC technique are discussed in details. Finally, the further prospects and challenges are outlined.

\section{Theory}

\subsection{Compound-tunable embedding potential}

The detailed description of the CTEP approach can be found in~\cite{Lomachuk:20_pccp,Maltsev:21_prb,Shakhova:22}; here we only recall the main points necessary to introduce the model of ``minimal cluster'' with broken covalent bonds. The general idea of the CTEP method is as follows. If one is interested in some localized processes or local properties of a crystal~\cite{Titov:14, Skripnikov:15b, Oleynichenko:18, Lomachuk:18en}, like X-ray emission spectra chemical shifts~\cite{Lomachuk:13, Shakhova:17rad, Shakhova:18en, Lomachuk:18en},
 parameters of hyperfine structure or effective Hamiltonians to search for the New physics beyond the Standard model (see \cite{Petrov:14, Petrov:23_HfF+_revisited} and references) or electronic excitations on $d$-, $f$-element impurity centers, one does not need to consider an infinite crystal as a whole. It is enough to choose only the central atom (typically a heavy metal ion) and its nearest ligands (typically anions). If only anions from the first coordination sphere are considered within the main cluster together with a central metal atom (cation) such a main cluster is called minimal.

Depending on the crystal structure, ligands can be either simple atomic anions or covalently bonded groups, such as the PO$_4^{3-}$ group in xenotime. In the latter case, the whole group is added to the main cluster to avoid breaking these essentially covalent bonds, thus significantly increasing its size. Such a main cluster is considered as an ``extended'' one. One of the primary goals of this work is to reduce the main cluster to a ``minimal'' one by including only those atoms of molecular-type anions which are direct neighbors of the central atom. Then such a ``minimal'' cluster can be studied using high-quality basis sets and small-core shape-consistent relativistic pseudopotentials for heavy atoms implying the explicit treatment of outercore atomic shells; the most accurate molecular correlation methods can be efficiently applied. The layer of cations next to the main cluster ligands comprises the nearest cationic environment (NCE). These atoms typically do not contribute explicitly to the properties of our interest which are localized on a main cluster, but their fractional charges compensate the overall negative charge of the main cluster and are necessary to reproduce the correct behaviour of a wavefunction within it.
The nearest anionic environment (NAE) is introduced analogously and includes anions from the layer next to the NCE ions. Moreover, within the CTEP approach \textit{all} electrons of the NCE ions are replaced with special compound-tunable pseudopotentials (CTPPs), which simulate the local (not only Coulomb but also exchange) response properties of environment near the main-cluster boundary. Such ions are further referred to as pseudoatoms. The more distant NAE ions are represented by negative charges or, optionally, CTPPs. Finally, the main cluster plus NCE and NAE can be optionally wrapped in a coat of fractional point charges, which together with the NCE plus NAE potentials form the electric field from environmental atoms of the crystal acting at the main-cluster atoms. If needed, the generalization for multi-center clusters (i.~e., those having a few central metal atoms) is straightforward.

\begin{figure*}
\includegraphics[width=\textwidth]{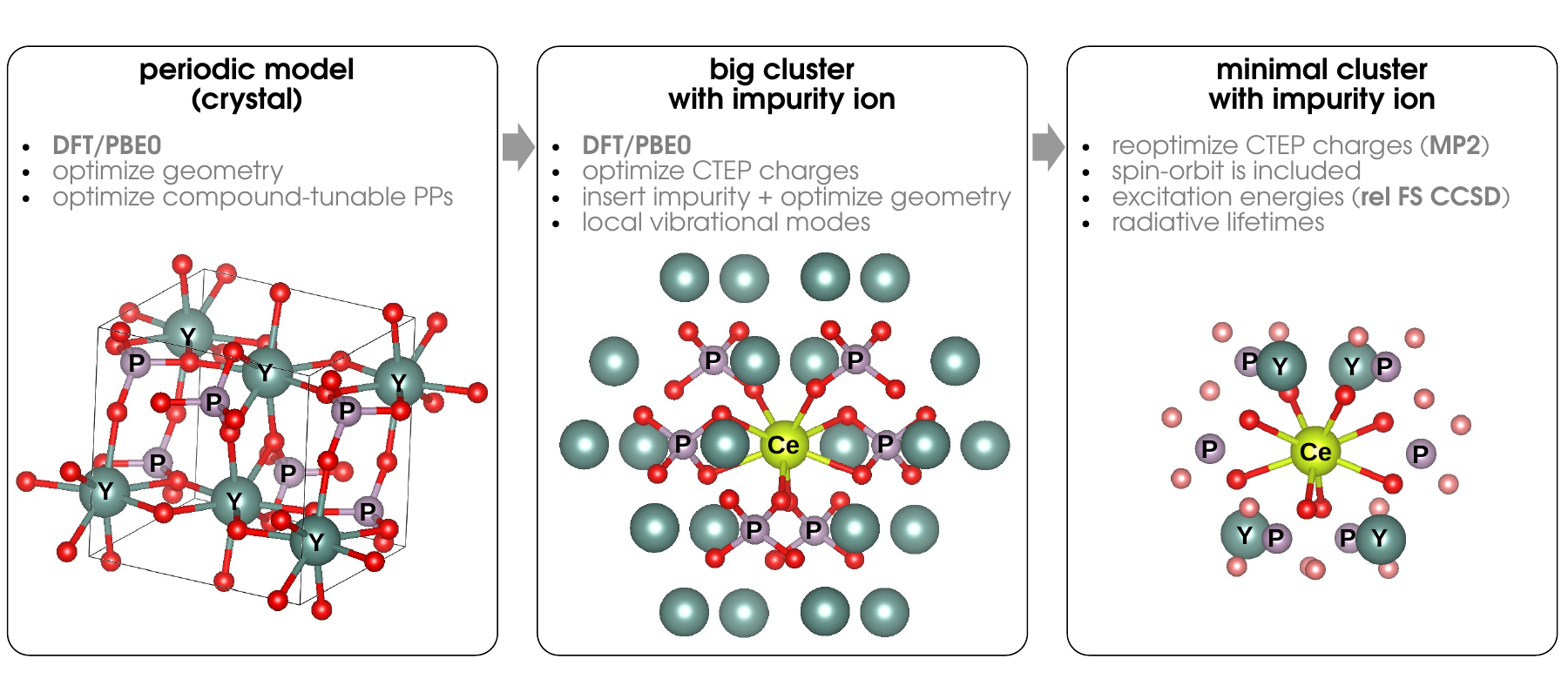}
\caption{General scheme of construction of a compound-tunable embedding potential for a minimal cluster model exemplified by the Ce-doped xenotime. The oxygen pseudoatoms of the anionic layer in the extended cluster model and CTEP partial charges are not shown for clarity. In both cluster models chemical bonds are shown for the main cluster areas. Cerium, yttrium, phosphorus and oxygen atoms are shown yellow, green, light-purple and red, respectively; for the minimal cluster oxygen pseudoatoms are shown in faded red.}
\label{fig:ctep-scheme}
\end{figure*}

The CTEP can be formally represented as a one-electron operator consisting of compound-tunable pseudopotentials $\hat{U}^{\rm CTPP}$ and/or fractional point charges $q_n$, centered at NCE and NAE pseudoatoms~\cite{Lomachuk:20_pccp,Shakhova:22} (the external layer of charges can be considered as a part of either NCE or NAE):
\begin{equation}
\hat{U}_{\rm CTEP}(\bm{r}) = 
\sum_{\substack{n\in {\rm NCE} \\ \;\;\cup {\rm NAE}}}
\left( \hat{U}^{\rm CTPP}_n(\bm{r} - \bm{r_n}) - \frac{q_n}{|\bm{r} - \bm{r}_n|} \right), \\
\label{eq:def-ctep}
\end{equation}
where $\bm{r}_n$ stands for the coordinates of the $n$-th pseudoatom.
Compound-tunable pseudopotentials $\hat{U}^{\rm CTPP}_n$
are essentially semilocal scalar-relativistic PPs 
designed in such a manner that the corresponding ``pseudo-ions'' have no occupied  electronic states. 
The general expression for the $\hat{U}^{\rm CTPP}(\bm{r})$ operator is 
\begin{align}
\hat{U}_n^{\rm CTPP}(\bm{r}) &= U_{n,L_n}(|\bm{r} - \bm{r_n}|) + \sum\limits_{l=0}^{L_n-1} U_{n,l}(|\bm{r} - \bm{r_n}|)P_{n,l} 
\nonumber \\
U_{n,l}(r) &= -\sum\limits_i d_{i,n,l}\cdot r^{p_{i,n,l}-2}\cdot \exp({-\zeta_{i,n, l} r^2}),
\label{eq:def-cttp}
\end{align}
where $P_{n,l}$ projects onto the subspace of functions with angular momentum $l$ with respect to the nuclear position $\bm{r}_{n}$. 
The coefficients $d_{i,n,l}$ and exponential parameters $\zeta_{i,n,l}$ are considered as adjustable parameter dependent only on the pseudoatom type, but not on its position.
The embedding potential defined by the formula~(\ref{eq:def-ctep}) can be readily used with almost any modern quantum chemistry software.
CTPP parameters and point charges defining the CTEP operator have to be calculated for each particular system under consideration; the general scheme of the CTEP construction is outlined in Fig.~\ref{fig:ctep-scheme}. Briefly, the main steps are:

(1) optimization of lattice parameters and atomic positions using periodic DFT calculations of a crystal without point defects;

(2) atoms which are planned to comprise the NCE layers are independently replaced by CTPPs. Their
parameters of Gaussian exponents
and coefficients are optimized at fixed geometry by minimizing the root mean square (RMS) force $|f|$ acting on atoms of the unit cell and defined as:
\begin{equation}
|f| = \sqrt{ \sum_{i=1}^N (\nabla_i E)^2 / N },
\label{eq:def-grad}
\end{equation}
where $N$ is the number of atoms in the main cluster and $\nabla_i E$ stands for the energy gradient with respect to the $i$-th atom coordinates;

(3) the cluster model for a fragment of interest is cut from the crystal structure; the geometry parameters stay fixed. Initial partial charges at NCE, NAE pseudoatoms and point nuclei beyond them are obtained from some rough approximation and then optimized by minimizing the RMS force on atoms of the main cluster. The electroneutrality of the whole cluster model must be maintained. At this point the cluster model of a pure crystal is constructed.

\subsection{Minimal cluster model of an impurity center}

To introduce substitute (or, in principle, admixture) ions (vacancies) one should place them into the cluster model instead of the main-cluster atoms and then re-optimize the cluster geometry. CTEP simulates the interactions with the rest of a pure crystal and must be fixed at this stage. An impurity atom(s) can distort local symmetry of its site. The resulting deformed extended cluster can be readily used to study properties induced by the presence of impurity atoms or/and vacancies.

The described scheme yields a cluster model which is computationally tractable at the FS~RCC level only if the main cluster area contains $\sim$10 atoms and $\sim$10$^2$ correlated electrons. These conditions are easily met for binary compounds (like ytterbium halides discussed in~\cite{Shakhova:22}), but fail 
for materials with more complex anions (e.\ g., phosphates). For example, for the model of xenotime Y(PO$_4$)$_6$@CTEP$_{\rm ext}$ from Ref.~\cite{Lomachuk:20_pccp} the main cluster area includes 31 atom. The obvious idea is to pass somehow to the minimal cluster model, for which only the atoms adjacent to the central one are included into the main cluster and treated at the highest possible level of theory, whereas the remaining atoms of complex anions are again replaced by CTEP. This approach would obviously worsen the description of chemical bonds inside the anion, but it would allow one to reduce the size of the problem for further correlation calculations. To achieve this goal, we propose to complete the basic CTEP generation scheme given above with the following step. The NCE and NAE layers are re-defined to include some atoms which previously belonged to the main cluster area; the layer of point charges (if presented) is re-defined consistently. The CTPP parameters are again inherited from the periodic calculations (see the step (2) above) or taken from calculations with some auxiliary cluster models, and point charges are optimized by minimizing the RMS force (Eq.~(\ref{eq:def-grad})) acting on the atoms of the \textit{new} (reduced) main cluster area. If a minimal cluster model is planned to be further treated at the CC level of theory, it looks logical to use some wave-function-based method (at least like MP2) to optimize point charges.

In principle one can swap the steps (4) and (5) and optimize the geometry of an impurity site for a minimal cluster model directly. However, this order of steps leads to significant decrease in accuracy of calculated geometry parameters and local properties, thus it cannot be recommended for practical use.

\subsection{Relativistic Fock space coupled cluster theory}

In the present study local electronic excitations on impurity $f$-element ions were simulated using the FS RCC theory~\cite{Kaldor:91,Visscher:01}. The method has been proved to be very efficient for systems with up to three electrons over the closed shell~\cite{Oleynichenko:CCSDT:20,Skripnikov:Bi:21,Eliav:Review:22}. For the CCSD approximation (with the single and double excitation operators included) errors in excitation energies typically do not exceed 0.05~eV~\cite{Eliav:02,Eliav:15}. Being a multi-state method fully compatible with relativistic Hamiltonians, FS RCC seems to be one of the best tools to simulate spectra of heavy-element compounds, including difficult cases with very dense electronic spectra and numerous nearly degenerate states.

In the particular case of the Ce$^{3+}$ and Th$^{3+}$ impurity ions one should start from the closed-shell ground state of the X$^{4+}$O$_8$@CTEP$_{\rm min}$ cluster (X = Ce, Th) considered as the Fermi vacuum (the $0h0p$ Fock space sector, i.~e. no particles and no holes with respect to the chosen vacuum) and then add one electron to low-energy (``active'') virtual spinors in all possible ways. The resulting model determinants thus belong to the $0h1p$ sector (zero holes, one particle). The target wavefunctions $\psi_i$ are obtained by the action of the normal-ordered exponentially parameterized wave operator on the model vectors $\tilde{\psi}_i$:
\begin{equation}
\ket{\psi_i} = \{e^T\} \ket{\tilde{\psi}_i}, \quad\quad \ket{\tilde{\psi}_i} = \sum_m C_{im} \ket{\Phi_m},
\label{eq:fscc-ansatz}
\end{equation}
where $T$ stands for the cluster operator, $\Phi_m$ are model determinants, and curly braces denote normal ordering. Within the CCSD approximation the cluster operator for the states with a single electron outside of the closed shell includes the following one- and two-body terms:
\begin{equation}
T = T^{0h0p}_1 + T^{0h0p}_2 + T^{0h1p}_1 + T^{0h1p}_2,
\label{eq:def-t-oper}
\end{equation}
where the subscripts indicate the excitation rank. Note that Fock space CC for the $0h1p$ sector considered here is nearly equivalent to the equation of motion (EOM) CC approach~\cite{Musial:08}.

To calculate transition properties within the FS RCC framework, 
the finite-order~\cite{Zaitsevskii:ThO:23} technique was developed.
Within the finite-order scheme an effective property operator matrix is expanded in powers of cluster amplitudes, and only linear and quadratic terms are retained in the present implementation. The resulting expression is constructed in a fully connected manner thus ensuring the size-consistency of transition moments calculated as off-diagonal matrix elements of the effective operator between the eigenvectors of the FS RCC effective Hamiltonian. The latter feature is of particular importance for systems with large number of electrons considered here. This method was applied in the present work to evaluate transition dipoles and then to estimate radiative lifetimes of excited $d$-states of impurity ions within the CTEP cluster model, assuming the vertical transition approximation.

The use of minimal cluster models allows one to use well-established and computationally efficient molecular codes to perform CC calculations. However, it should be pointed out that an impurity atom can reduce symmetry of a site. In the specific case of the doped xenotime the symmetry of both doped minimal clusters after geometry optimization was reduced to $C_1$. If the relativistic Hamiltonian (e.~g. two-component pseudopotential) is used one has to perform all calculations in complex arithmetic and no computational savings due to symmetry are possible. These circumstances require the CC code to be very efficient and oriented at parallel execution in order to get results in a reasonable time.

\section{Computational details}

\begin{figure}[!h]
\centering  
\includegraphics[width=0.8\columnwidth]{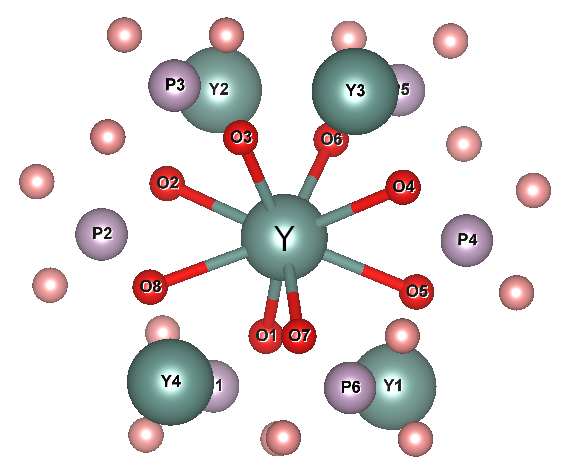} 
\caption{Minimal xenotime cluster YO$_8$@CTEP$_{\rm min}$. Chemical bonds are depicted between atoms of the main cluster [YO$_8$]. The NCE consists of yttrium and phosphorus pseudoatoms, the NAE consists of oxygen pseudoatoms (shown in the faded red color).}
\label{fig:cluster}
\end{figure}

In the present study the minimal cluster of xenotime YO$_8$@CTEP$_{\rm min}$ (see Fig.~\ref{fig:cluster}) and clusters with Ce$^{3+}$ and Th$^{3+}$ impurity ions substituting the central Y$^{3+}$, further denoted as CeO$_8$@CTEP$_{\rm min}$ and ThO$_8$@CTEP$_{\rm min}$, respectively, were constructed according to the technique described in the previous section. The cluster models employed consist of
\sli1 central yttrium (or impurity) atom and 8 nearest oxygen atoms treated explicitly the electronic structure calculations (the main cluster);
\sli2 6 phosphorus pseudoatoms and 4 yttrium pseudoatoms in the nearest cationic environment (NCE);
\sli3 16 oxygen pseudoatoms in the nearest anionic environment (NAE);
\sli4 9 point charges that improve the modeling of the rest of a crystal and restore electroneutrality for the whole cluster model. Positions of atomic nuclei for all cluster models described in the present paper were taken from Ref.~\cite{Lomachuk:20_pccp} where they were computed for the extended cluster model. The minimal cluster model YO$_8$@CTEP$_{\rm min}$ is not reliable for geometry optimizations of crystal fragments with impurities. Thus they were performed for extended cluster models which were then cut to the minimal ones to perform their coupled cluster studies as it was described in Section II.A. The geometries of all cluster models constructed in the paper together with corresponding CTEP parameters are available in Supplementary Material.

The {\sc NWChem} program package~\cite{NWCHEM} was used to construct cluster models, optimize CTEP parameters and perform auxiliary time-dependent DFT (TD-DFT) calculations of excitation spectra used to estimate
corrections due to reduction of the cluster model size.
%
The PBE0 hybrid exchange-correlation functional~\cite{Adamo:99} was used in DFT calculations throughout the paper (unless otherwise specified), the spin-unrestricted DFT was used for open-shell species. CTEP parameters for the extended Y(PO$_4$)$_6$@CTEP$_{\rm ext}$ cluster were taken from our previous paper~\cite{Lomachuk:20_pccp}.
Due to the numerical instabilities of the conventional TD-DFT method,
the
Tamm-Dancoff approximation~\cite{Jensen:17_IntoCompCh}
was used throughout the paper. It should be pointed out that the widely used TD-DFT method, which allows one to treat large systems~\cite{Cramer:09, Escudero:17, Sun:21, Liang:22}, can deal with one-electron excitation processes only and gives an unacceptably low accuracy with typical errors $\sim$1~eV in a large number of practical cases. Thus it cannot serve as a basis for really precise modeling.

Basis sets and pseudopotentials used in DFT calculations were as follows. For the central atoms scalar-relativistic versions of the shape-consistent relativistic pseudopotentials for Y, Ce (28 electrons in core) and Th (60 electrons in core) were employed~\cite{Mosyagin:10a,Mosyagin:16,Mosyagin:17}. Primitive Gaussian basis set ($5s5p3d$) was used for Y, while basis sets for Ce and Th were more extended and comprised ($10s6p6d7f$)/[$5s4p6d7f$] (Ce) and ($7s7p6d4f$)/[$5s6p4d3f$] (Th) functions. For oxygen atoms in the main cluster area we used the all-electron ($10s6p1d$)/[$4s3p1d$] basis set from~\cite{Peintinger:13}. Within the minimal cluster models the P, Y, and O atoms in the NCE and NAE layers were simulated by the pseudopotentials with compound-tunable parameters describing interactions with all the electrons attributed to an atomic ion in accord to its formal oxidation state
(in other words, such an ion was treater without electrons explicitly associated with it, 0ve-PPs)

Basis sets and pseudopotentials for the extended cluster models were taken from~\cite{Lomachuk:20_pccp}. To fit the formal oxidation states of main cluster atoms
and saturate chemical bonds between atoms of main cluster and environment
13 electrons were added to the main cluster area. Thus, for the minimal xenotime cluster with the central Y$^{3+}$ ion the total number of the explicitly treated electrons was equal to 88. The charge of these additional electrons was compensated by introducing fractional charges at pseudoatoms of the NCE and NAE layers (see Table~\ref{table:charges}). These charges along with the parameters of the 0ve-PPs were adjusted to minimize average force acting on atoms of the main cluster area (Eq.~(\ref{eq:def-grad})).
The CTPPs for the P, Y and O pseudoatoms were assumed to have the form (\ref{eq:def-cttp}) with small number of terms in expansions. 
The $d_{i,n,l}$ and $\zeta_{i,n,l}$ coefficients in Eq.~(\ref{eq:def-cttp}) were considered as variable parameters except for the terms with $r^{-2}$ ($p_{i,n,l} = 0$); in this special case only the exponential parameter $\zeta_{i,n,l}$ was optimized, while the coefficient $d_{i,n,l}$ was taken equal to $-l(l+1)/2+L(L+1)/2$, see Eq.~(\ref{eq:def-cttp}) for the definition of $L$.
To describe properly the polarization of one-electron states corresponding to the P--O and Y--O bonds between the main cluster and NCE, the Y and P pseudoatoms from NCE were supplemented with minimal basis sets of type $(1s1p1d)$.


\begin{table}[!ht]
\caption{Fractional charges (a.~u.) on pseudoatoms of the nearest cationic and anionic environments of xenotime minimal cluster models with central X = Y$^{3+}$, Ce$^{3+}$, Th$^{3+}$ ions}
\label{table:charges}
\renewcommand{\arraystretch}{1.3}
\begin{tabular*}{\columnwidth}{l@{\extracolsep{\fill}}ccc}
\hline
\hline
& YO$_8$@CTEP$_{\rm min}$ & CeO$_8$@CTEP$_{\rm min}$ & ThO$_8$@CTEP$_{\rm min}$ \\
\hline
Y       &  +0.41 & +0.72 & +1.22 \\
[2ex] P\fn1   &  +4.01 & +4.66 $\pm$ 0.07 & +4.38 $\pm$ 0.02 \\
              &  +3.90 & +3.78 $\pm$ 0.13 & +4.06 $\pm$ 0.03 \\ 
[2ex] O\fn1   &  $-$0.21 &   +0.05 $\pm$ 0.05 & $-$0.29 $\pm$ 0.03 \\
              &  $-$0.50 & $-$1.05 $\pm$ 0.10 & $-$0.97 $\pm$ 0.02 \\ 
              &  $-$1.64 & $-$1.27 $\pm$ 0.05 & $-$1.28 $\pm$ 0.01 \\ 
              &          & $-$1.40 $\pm$ 0.05 & $-$1.34 $\pm$ 0.01 \\ 
[2ex] $f$, a.~u.\fn2 &   $10^{-5}$      & $1.6 \cdot 10^{-3}$ & $5 \cdot 10^{-3}$ \\
\hline
\hline
\end{tabular*}
\footnotetext[1]{The dispersion of fractional charges of NCE and NAE pseudoatoms reflects the local symmetry lowering for the clusters with admixture ions. For the ``pure'' YPO$_4$ cluster two kinds of the PO$_4$ groups exist, with either one or two oxygen atoms adjacent to the central ion, resulting in the two and three groups of charges for P and O pseudoatoms, respectively. The clusters with the Ce$^{3+}$ and Th$^{3+}$ admixture ions are more distorted.
}
\footnotetext[2] {Average of the absolute value of the force acting on the atoms of the main cluster area (see Eq.~(\ref{eq:def-grad})).
}
\end{table}

Relativistic Fock space coupled cluster calculations of excitation energies were performed with the semilocal versions of the shape-consistent generalized relativistic pseudopotentials substituting 28 and 60 core electrons of Ce~\cite{Mosyagin:17} and Th~\cite{Mosyagin:16,Zaitsevskii:ThO:23}, respectively. Highly accurate description of $d$- and especially $f$-shells requires the use of correlation-consistent basis set with high angular momenta functions for a lanthanide or actinide atom. The set of primitive Gaussians was loosely based on the Dyall's quadruple-zeta quality set~\cite{Dyall:07,Gomes:10}. Contracted basis functions for Ce were obtained as average natural orbitals (ANO~\cite{Widmark:90}) by averaging analytic density matrices calculated at the single-reference relativistic CCSD level~\cite{AnalyticDM:22} for the ground state of Ce$^{4+}$ and the $5d_{3/2}$, $5d_{5/2}$, $4f_{5/2}$, $4f_{7/2}$ states of Ce$^{3+}$. The final version of the basis set was quite compact and comprised ($15s14p11d13f7g6h$)/[$3s4p4d3f2g1h$] functions. Contracted ANO-type basis set for Th was taken from~\cite{Zaitsevskii:ThO:23} and restricted to ($11s10p9d8f7g6h$)/[$4s4p4d3f2g1h$].

To estimate the quality of the basis sets employed, 
FS RCCSD calculations of energy levels of the Ce$^{3+}$ and Th$^{3+}$ ions with both contracted and uncontracted basis sets were performed (see Tables~\ref{tab:ce-basis} and~\ref{tab:th-basis}). One can immediately see that $4s4p$-shells of Ce and $5s5p$-shells of Th contribute up to $0.34$~eV and 0.13~eV to excitation energies, respectively, and thus were correlated in all further calculations. The excellent agreement of excitation energies obtained within the CCSD approximation and exhaustive basis sets including $i$-functions with their experimental counterparts is not surprising (actually the same result was also reported in~\cite{Eliav:02}). For these charged systems, which can be effectively considered as the one-electron ones, contributions of triple excitations are negligible (of order $\sim0.01$~eV), while contributions from non-local terms in relativistic pseudopotentials and basis functions with angular momenta higher than $i$ which were missed here are of order $\sim0.05$~eV, but compensate each other. The deficiencies of contracted basis sets are partially overcome in solid-state calculations due to basis functions centered on atoms adjacent to the central one. The final upper bound estimate for the errors due to the use of semilocal relativistic PPs and contracted basis sets is 0.2~eV for Ce and 0.1~eV for Th.

\begin{table}
\caption{Deviations of excitation energies (in eV) of the Ce$^{3+}$ atomic ion (from the ground state $4f$ $^2F^o_{5/2}$) calculated at the FS RCCSD($0h1p$) level with different basis sets from the experimental data~\cite{Sansonetti:05}.}
\label{tab:ce-basis}
\renewcommand{\arraystretch}{1.3}
\begin{tabular*}{\columnwidth}{l@{\extracolsep{\fill}}ccccc}
\hline
\hline
 & & \multicolumn{2}{c}{uncontracted basis}  & \multicolumn{2}{c}{contracted basis} \\
 & & \multicolumn{2}{c}{($15s14p11d13f7g6h5i$)}  & \multicolumn{2}{c}{[$3s4p4d3f2g1h]$} \\
 \cmidrule{3-4} \cmidrule{5-6}
Term & Exptl & $4s4p$ corr & $4s4p$ frozen & Ce$^{3+}\footnotemark[1]$ & Ce$^{3+}$+8O\footnotemark[2] \\
\hline
$^2F^o_{7/2}$ & 0.28 & 0.00 & $-$0.01 & $-$0.01 & $-$0.03 \\
$^2D_{3/2}$   & 6.17 & 0.00 & $-$0.34 & $-$0.13 & $-$0.17 \\
$^2D_{5/2}$   & 6.48 & 0.00 & $-$0.34 & $-$0.13 & $-$0.18 \\
\hline
\hline
\end{tabular*}
\footnotetext[1]{Isolated Ce$^{3+}$ ion, $4s4p$ shells were correlated.}
\footnotetext[2]{Ce$^{3+}$ ion plus 8 ghost atoms centered at positions of oxygens taken from the cluster model, basis set for oxygen [$4s3p1d$] was used for ghost atoms, $4s4p$ shells were correlated.}
\end{table}

\begin{table}
\caption{Deviations of excitation energies (eV) of the Th$^{3+}$ atomic ion (from the ground state $5f$ $^2F^o_{5/2}$) calculated at the FS RCCSD($0h1p$) level with different basis sets from the experimental data (see~\cite{Klinkenberg:49,Eliav:02} and references therein).}
\label{tab:th-basis}
\renewcommand{\arraystretch}{1.3}
\begin{tabular*}{\columnwidth}{l@{\extracolsep{\fill}}ccccc}
\hline
\hline
 & & \multicolumn{2}{c}{uncontracted basis}  & \multicolumn{2}{c}{contracted basis} \\
 & & \multicolumn{2}{c}{($11s10p9d8f7g6h5i$)}  & \multicolumn{2}{c}{[$4s4p4d3f2g1h$]} \\
 \cmidrule{3-4} \cmidrule{5-6}
Term & Exptl & $5s5p$ corr & $5s5p$ frozen & Th$^{3+}\footnotemark[1]$ & Th$^{3+}$+8O\footnotemark[2] \\
\hline
$^2F^o_{7/2}$ & 0.54 & 0.00 & $-$0.02 & 0.01 &    0.01 \\
$^2D_{3/2}$   & 1.15 & 0.00 & $-$0.13 & 0.07 &    0.00 \\
$^2D_{5/2}$   & 1.81 & 0.00 & $-$0.13 & 0.04 & $-$0.03 \\
$^2S_{1/2}$   & 2.82 & 0.00 & $-$0.11 & 0.20 &    0.06 \\
$^2P^o_{1/2}$ & 7.47 & 0.00 & $-$0.11 & 0.12 &    0.04 \\
$^2P^o_{3/2}$ & 9.06 & 0.00 & $-$0.12 & 0.02 & $-$0.07 \\
\hline
\hline
\end{tabular*}
\footnotetext[1]{Isolated Th$^{3+}$ ion, $5s5p$ shells were correlated.}
\footnotetext[2]{Th$^{3+}$ ion plus 8 ghost atoms centered at positions of oxygen atoms taken from the cluster model, basis set for oxygen [$4s3p1d$] was used for ghost atoms, $5s5p$ shells were correlated.}
\end{table}

For the Ce$^{3+}$ atomic ion excitations to the $6s$ $^2S_{1/2}$ (10.7~eV) and $6p$ (15.2~eV) states lie well above the band gap ($\sim8.6$~eV) and thus are not expected to be observed in solids. In contrast, in the case of the free Th$^{3+}$ ion these energy levels can lie below the band gap (see Table~\ref{tab:th-basis}). The situation is not quite clear for the $7p_{3/2}$ level ($\sim9.1$~eV) and deserves a detailed discussion since in principle this level can be shifted down in a solid matrix. Anyway, the $7s$ and $7p$ states are expected to be rather diffuse, thus for these states the minimal cluster model may fail. According to these considerations, active space in FS RCC calculations comprised spinors which can be regarded as the $4f$, $5d$ (Ce) and $5f$, $6d$, $7s$, $7p$ (Th) spinors of an impurity ion split in the crystal field. The model with single and double excitations (CCSD) was employed, and $1s$-shells of oxygen atoms were kept frozen in CC calculations. Virtual energy cutoff was taken to be +50 a.u. Transition dipole moments needed to estimate radiative lifetimes were evaluated using the finite-order approach presented in~\cite{Zaitsevskii:ThO:23} (including terms up to quadratic in $T$).

Molecular integrals for coupled cluster calculations were obtained using the DIRAC19 software~\cite{DIRAC_code:19,Saue:20}, and subsequent FS RCCSD calculations of energies and transition dipole moment matrix elements were carried out using the EXP-T program package~\cite{Oleynichenko:EXPT:20,EXPT:23}. Crystal and molecular structures were visualized using the VESTA software~\cite{Momma:11}. The code for optimization of CTEP parameters uses the SciPy library~\cite{Virtanen:20}.

\section{Results and discussion}

Optimized X--O (X = Y, Ce, Th) and P--O bond lengths obtained for the minimal cluster models under consideration are listed in Table~\ref{table:bonds}. For the pure xenotime there are only two different kinds of oxygen atoms with different Y--O bond length values (see Table~\ref{table:bonds} for details). Since the clusters with the Ce$^{3+}$ and Th$^{3+}$ impurity ions were prepared with the geometries taken from preliminary optimized extended clusters Ce(PO$_4$)$_6$@CTEP$_{\rm ext}$ and Th(PO$_4$)$_6$@CTEP$_{\rm ext}$, and not from the periodic calculations as for YPO$_4$, the symmetry of the metal site of these clusters is significantly lower than that for the pure xenotime ($D_{2d}$). This leads to more complicated structures of CTEPs for the cluster models of YPO$_4$:Ce$^{3+}$ and YPO$_4$:Th$^{3+}$ (see the fractional charges in Table~\ref{table:charges}); all the CTEP charges had to be optimized separately. As a result, for the clusters with impurity centers we were not able to reduce the RMS forces to nearly zero ($10^{-5}$ a.~u. as for the model of pure YPO$_4$, see Table~\ref{table:charges}), though the embedding uncertainties ${\sim}{10}^{-3}$ a.~u. attained for our clusters with Ce and Th impurities are also lying within the uncertainties of periodic structure medium-core PP/DFT models \cite{Maltsev:21_prb, Shakhova:22} containing Ce and Th atoms.

\begin{figure}
\begin{tabular}{c}
\includegraphics[width=\columnwidth]{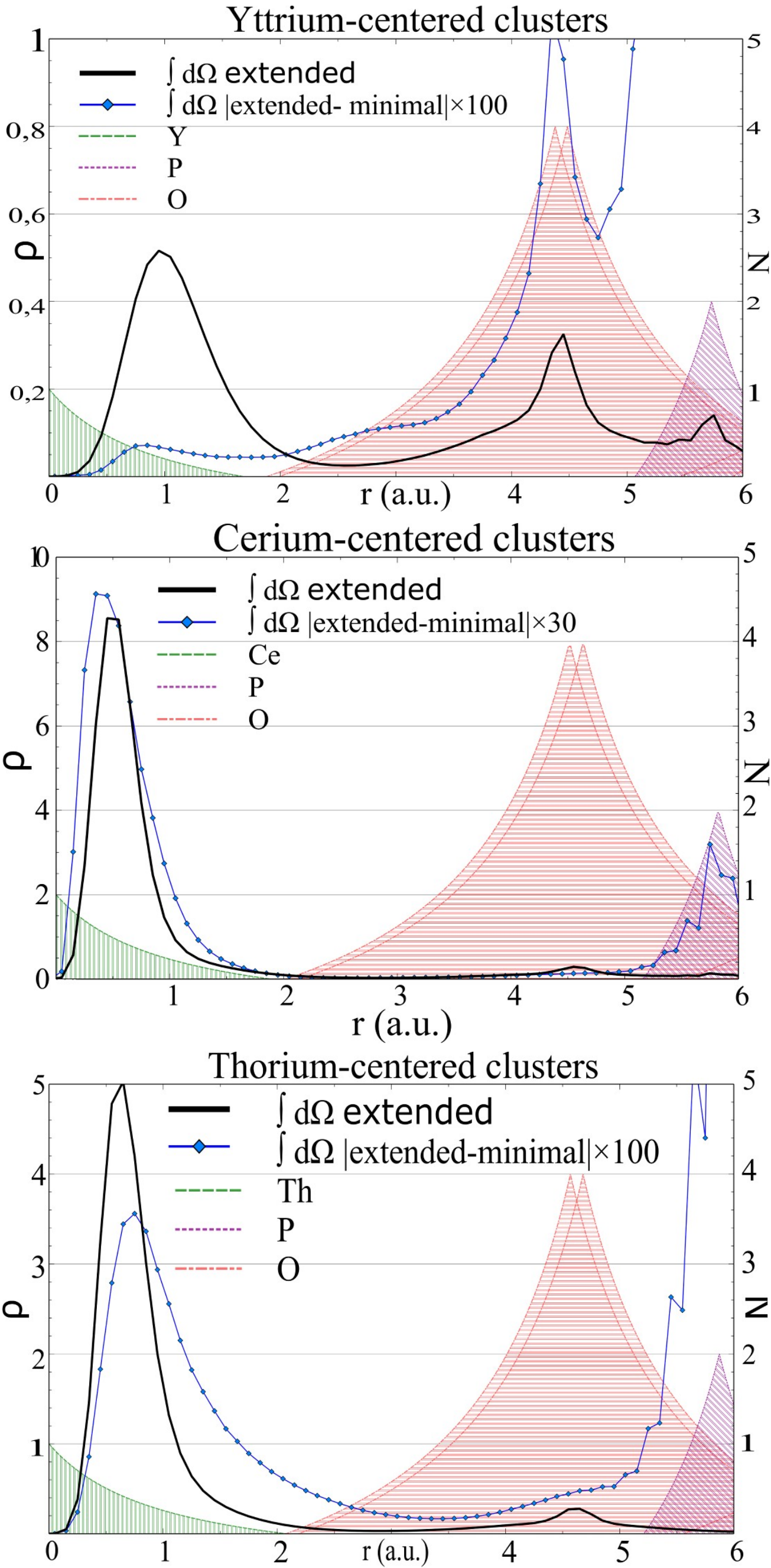}
\end{tabular}
\caption{
The radial dependence of electronic density differences for the studied extended and minimal clusters (see Eq.~(\ref{eq:density}) and text below the formula for details). Black lines represent total densities.
Due to the using of pseudopotentials for the central yttrium, cerium, and thorium ions, the radial total density maximums, associated with this ions are shifted from the point $r=0.0$. The height of the maximums for Y, Ce, and Th ions correspond to the number of their electrons treated explicitly. 
Colored solid lines correspond to the integral of absolute difference $\rho(r)$ multiplied by factor of 100 (30 in the  case of cerium ion). Filled peaks in the background qualitatively represent positions of the neighbour atoms (color denotes an atom type, width at the bottom is equal to crystal radius, and the peak height is proportional to the number of atoms (right vertical axis) at the
same distance from the center).}
\label{fig:densities}
\end{figure}

\begin{table}[!h]
\caption{Bond lengths (in \AA) in minimal xenotime clusters with central X = Y$^{3+}$, Ce$^{3+}$, Th$^{3+}$ ions.}
\label{table:bonds}
\renewcommand{\arraystretch}{1.3}
\begin{tabular*}{\columnwidth}{l@{\extracolsep{\fill}}ccc}
\hline
\hline
& YO$_8$@CTEP$_{\rm min}$ & CeO$_8$@CTEP$_{\rm min}$ & ThO$_8$@CTEP$_{\rm min}$ \\
\hline
P--O\fn1   &  1.566   &  1.562    & 1.564 \\
           &          &  1.567    & 1.567 \\
[2ex]
X--O\fn1  &  2.317   &  2.389    & 2.418 \\
          &  2.371   &  2.451    & 2.476 \\
[2ex]
O--O\fn1 &  2.421   & 2.32 $\pm$ 0.01 & 2.44 $\pm$ 0.03 \\ 
           &          & 2.459           & 2.61 $\pm$ 0.01 \\
\hline
\hline
\end{tabular*}
\footnotetext[1]{Bond lengths which differ by no more than 0.001~\AA~from the listed ones for the highly symmetric YO$_8$@CTEP$_{\rm min}$ cluster are not shown. For the Ce- and Th-containing clusters symmetry is significantly lower and bond length values are spread within some intervals.}
\end{table}
To estimate the quality of the minimal cluster models, electronic density cube files were obtained for the extended and corresponding minimal cluster models. The cube grid was chosen to be the same for all cases with the orthogonal unit vectors of about 0.07 a.u. As a quantitative criterion we provide the difference between the minimal and extended cluster densities calculated using the following formula:
\begin{equation}
    \label{eq:density}
    \rho(r) = \frac{1}{4\pi} \int d\Omega\ |\rho_{ext}(\vec{r})-\rho_{min}(\vec{r})|.
    \end{equation}
In the equation above, the $\rho_{ext}(\vec{r})$ and $\rho_{min}(\vec{r})$ are the electronic densities in the extended and minimal cluster models, respectively; the coordinate origin is located at the central atom position and integration is performed over angular variables. In Fig.~\ref{fig:densities} this value is plotted for all the clusters under study. The dashed peaks in the background qualitatively represent the electronic density of atoms from the main cluster. The black curve is the total density, and the difference curves for the clusters with yttrium and thorium central ions are multiplied by factor 100, so that intersections between total and difference curves correspond to the 1\% deviation, while the multiplier value for the YPO$_4$:Ce$^{3+}$ clusters equals to 30. At the distance $r = 4 - 5$ a.~u., which corresponds to the oxygen atoms positions, the difference increases by about 3\% for all considered clusters. This is due to the fact that oxygen electronic states slightly differ for extended and minimal cluster models. At the distance about $r=5.5$ a.~u.\ from the central atom this value increases more rapidly, marking the border of the main cluster.

\begin{table}
\caption{Band gap estimates for pure xenotime obtained within periodic structure (crystal) and the cluster model calculations. $|f|$ stands for the RMS force per atom in the main cluster area (Eq.~(\ref{eq:def-grad})).}
\label{table:bandgap}
\renewcommand{\arraystretch}{1.3}
\begin{tabular*}{\columnwidth}{l@{\extracolsep{\fill}}cc}
\hline
\hline
Method/cluster model          &  Band gap, eV      & $|f|$, a.u. \\
\hline 
Exptl.~\cite{Wang:09}                         & 8.6 & -- \\
\multicolumn{3}{c}{\it periodic calculations} \\
PBE0/crystal exp. geometry\footnotemark[1]    & 8.8 & -- \\
PBE0/crystal\footnotemark[1]                  & 8.5 & -- \\
PBE96/plane wave\footnotemark[2]                      & 5.8 & -- \\
%
\multicolumn{3}{c}{\it TD-DFT calculations\footnotemark[3]} \\
%
PBE0/extended cluster   & 8.5    & $10^{-5}$ \\
%
\multicolumn{3}{c}{\it E(triplet) - E(singlet)\footnotemark[4]} \\
%
PBE0/extended cluster   & 9.0    & $10^{-5}$ \\
[2ex]
%
\hline
\hline
\end{tabular*}
\footnotetext[1]{Indirect band gap from crystal structure calculations, performed within the DFT/PBE0 framework using the {\sc{Crystal17}} \cite{Dovesi:18} software.}
\footnotetext[2]{Plane-wave DFT calculations with the PBE96 functional. Data from the Materials Project~\cite{Jain:13,Munro:20}.}
\footnotetext[3]{Geometry was optimized for ground state.}
\footnotetext[4]{The band gap value is estimated as a difference between full energies of the ground (singlet) and first excited (triplet) states, for both states geometry optimization of the main cluster atomic positions was performed.}
\footnotetext[5]{Single-point energy calculations without geometry optimization.}
%
\end{table}

The estimates for the band gap values in xenotime are provided in Table~\ref{table:bandgap}. Note that the estimates obtained for cluster models should be discussed with caution since the band structure cannot be correctly defined for finite-size systems. However, to some extent lowest unoccupied molecular orbitals can be identified with the states lying at the bottom of the conduction band. As it can be seen from Table~\ref{table:bandgap}, while passing from fully periodic to the cluster model, the band gap estimate changes rather moderately. It is interesting that the generalized gradient approximation completely fails to predict the band gap value (5.8~eV for the PBE96 functional~\cite{Perdew:96}) while hybrid functionals PBE0 and HSE06~\cite{Heyd:03} reproduce the experimental datum quite well, thus indicating that the exact exchange contribution is important for the successful description of electronic structure of xenotime. Besides, as one can see from the table, the xenotime cluster of extended size provides the band gap in a rather good agreement with the corresponding periodic structure value. So, it looks like that our Y(PO$_4$)$_6$@CTEP$_{\rm ext}$ model can be used for further refining the theoretical band gap value within accurate WFT modeling of xenotime (that is in our future plans).

\begin{figure}[!h]
\centering
\includegraphics[width=0.85\columnwidth]{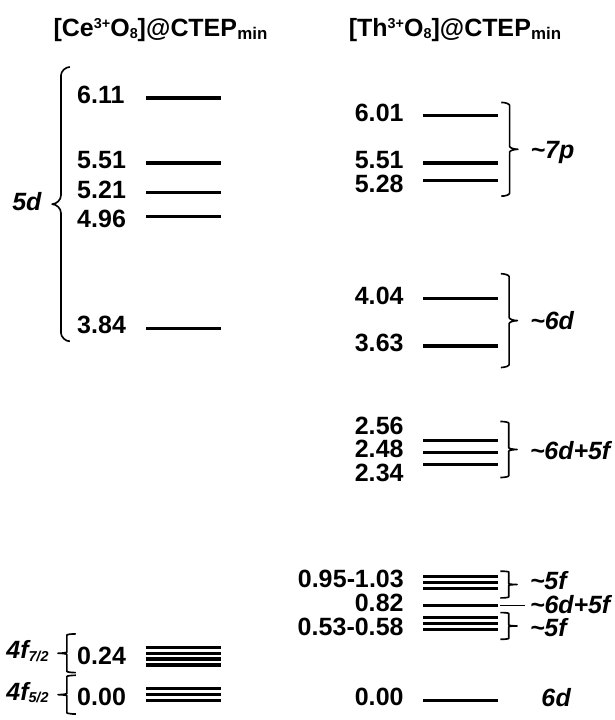}
\caption{Energy levels of impurity ions in xenotime matrix. For Ce$^{3+}$, the values estimated from experimental data~\cite{Dorenbos:13} are plotted. The levels for Th$^{3+}$ were estimated by the FS RCCSD calculation (see Table~\ref{tab:exc-fscc-th}).}
\label{fig:df_levels}
\end{figure}

We now turn to studying the local excitation processes in doped xenotime materials using minimal cluster models. The general pattern of electronic levels of an impurity ion is defined mainly by the two competing effects: spin-orbit interaction and crystal field splitting (see Fig.~\ref{fig:df_levels}). The Y$^{3+}$ site in xenotime possesses the $D_{2d}$ local symmetry, however, for impurity ions symmetry is further lowered to $C_1$ and one can expect that degeneracies of sublevels with different $m_j$ quantum numbers can be completely removed.

Experimentally measured low-temperature vibronic spectra of doped YPO$_4$:Ce$^{3+}$ crystals published in~\cite{Sytsma:93,Laroche:01,Pieterson:02, Krumpel:09} clearly demonstrate that for the $5d$-levels of Ce$^{3+}$ spin-orbit and crystal field effects are comparable, though with a predominance of the latter. For the $4f$-levels the splitting induced by the crystal field is nearly an order of magnitude smaller than the spin-orbit splitting (which is nearly equal to $0.24$~eV, cf.~0.28~eV for the Ce$^{3+}$ atomic ion) and cannot be directly observed in experiments since the vibronic bands corresponding to emission to different $4f_{m_j}$ sublevels strongly overlap with each other, and the spectrum cannot be further resolved. Such a modest influence of the crystal field is due to the fact that the $4f$-orbitals are quite compact and localized mainly in the inner region of the cerium ion, they are not mixed significantly with orbitals of ligands. 
Experimental spectra for thorium-doped xenotime crystals are unavailable, but one can expect a much more entangled picture since the $5f$-orbitals of Th$^{3+}$ are more diffuse than those of Ce$^{3+}$; the crystal field splitting for them would increase and interplay of the crystal field with spin-orbit interaction results in a quite entangled picture. Based on the experimental data on energy levels of the Th$^{3+}$ atomic ion (see Table~\ref{tab:th-basis}) one can also expect the presence of the $7s$ and $7p$ levels of the thorium impurity below the conduction band, and transitions to them from the ground $6d^1$ state can become allowed (and observable experimentally) due to the hybridization with orbitals of PO$_4$ groups.

\begin{table}
\caption{Calculated energy levels (eV) of the Ce$^{3+}$ impurity ion in xenotime matrix and lifetimes of excited $5d$-states.
Corrections due to reduction of the cluster model size
$\Delta$ are estimated at the TD-DFT level; the $\Delta$ values cannot be unambiguously extracted for the $4f$ states and thus are not given. Lifetimes of excited $4f$ states are of order of 50~ms and higher; they are not given since large uncertainties for them are expected. Experimentally measured lifetimes for the first $5d$ level are 12.5 ns~\cite{Lai:07} and 23 ns~\cite{Laroche:01}.
}
\label{tab:exc-fscc}
\renewcommand{\arraystretch}{1.3}
\begin{tabular*}{\columnwidth}{l@{\extracolsep{\fill}}ccccl}
\hline
\hline
 & \multicolumn{3}{c}{E, eV} & $\tau$, ns \\
\cmidrule{2-4} \cmidrule{5-5}
No. & CCSD & CCSD+$\Delta$ & Exptl\footnotemark[1] & CCSD & Composition \\
\hline
\multicolumn{6}{c}{\it 4f levels} \\
1 & 0.00 & -- & 0.00 & -- & $4f_{5/2}^{1.0}$ \\
2 & 0.06 & -- &      & -- & $4f_{5/2}^{0.9}$ \\
3 & 0.07 & -- &      & -- & $4f_{5/2}^{0.9}$ $4f_{7/2}^{0.1}$ \\
4 & 0.24 & -- & 0.24 & -- & $4f_{7/2}^{0.9}$ \\
5 & 0.26 & -- &      & -- & $4f_{7/2}^{0.9}$ \\
6 & 0.32 & -- &      & -- & $4f_{7/2}^{0.9}$ \\
7 & 0.33 & -- &      & -- & $4f_{7/2}^{0.9}$ \\
\multicolumn{6}{c}{\it 5d levels} \\
 8 & 3.43 & 3.58 & 3.72 (3.84) & 23.7 & $5d_{3/2}^{0.4}$ $5d_{5/2}^{0.5}$ \\
 9 & 5.12 & 4.99 & (4.96) & 10.6 & $5d_{3/2}^{0.5}$ $5d_{5/2}^{0.2}$ \\
10 & 5.26 & 5.12 & (5.21) & 11.5 & $5d_{3/2}^{0.2}$ $5d_{5/2}^{0.6}$ \\
11 & 5.85 & 5.20 & (5.51) & 10.4 & $5d_{3/2}^{0.2}$ $5d_{5/2}^{0.5}$ \\
12 & 6.34 & 5.90 & (6.11) &  7.7 & $5d_{3/2}^{0.2}$ $5d_{5/2}^{0.4}$ \\
\hline
\hline
\end{tabular*}
\footnotetext[1]{Experimental data presented in the work \cite{Krumpel:09}. 
Parenthesized values are taken from the review paper \cite{Dorenbos:13}.}
\end{table}

These qualitative considerations are generally reproduced in relativistic coupled cluster calculations; results for the Ce$^{3+}$ and Th$^{3+}$ clusters are presented in Tables~\ref{tab:exc-fscc} and~\ref{tab:exc-fscc-th}, respectively. The energy levels obtained in the FS RCCSD calculation for the minimal cluster, CeO$_8$@CTEP$_{\rm min}$, can be straightforwardly correlated with their experimentally measured counterparts. However, the accuracy of calculated energies is not perfect, the mean absolute deviation (MAD) for the $5d$ levels is 0.24~eV and the largest deviations were obtained for the first and fourth $5d$-levels ($-$0.41~eV and 0.34~eV, respectively). To confirm the composition of electronic states, the projection population analysis (see~\cite{Dubillard:06,Oleynichenko:18} and references therein) was applied to natural spinors corresponding to each of these states (see the last column in Table~\ref{tab:exc-fscc}); these spinors were obtained in an approximate manner by diagonalization of density matrices restricted to the model space. The obtained occupation numbers clearly demonstrate that the assumption about localized nature of these excitations on the Ce$^{3+}$ ion is satisfied with a good accuracy, maybe except for the last level with nearly 40\% of density of the unpaired electron leaving the impurity ion.

The deviations of calculated excitation energies from their experimental counterparts are mainly determined by two sources of error: the basis set incompleteness and the minimal size of the embedded cluster model used. The error introduced by the former factor for the Ce-doped cluster is expected to be not exceeding 0.2 eV (see Table~\ref{tab:ce-basis}). The significance of the cluster model
reduction
is not {\it a priori} obvious. It seems reasonable to estimate corrections
due to reduction of the cluster model size
(or at least to set some limitations for the corresponding error) using the TD-DFT approach, since such calculations are feasible for both minimal and extended clusters. The Kramers-unrestricted two-component TD-DFT model compatible with the use of fully relativistic
pseudopotentials is not available to date, thus the scalar-relativistic approximation has to be used. Note, however, that such an approximation is quite reasonable for lanthanides as impurities in the crystals containing light atoms since only the correction on extension of the minimal cluster is then used in our final results (see Eq.~(\ref{Eq.ext-min}) below), not absolute values of excitation energies. The TD-DFT calculations were performed for both extended and minimal cluster models. Errors of TD-DFT excitation energies for the $5d$-levels of Ce$^{3+}$ are in the range of $0.5-1.0$~eV, which is significantly worse than for FS RCCSD. However, the correction for the cluster model 
reduction
defined as
\begin{equation}
\Delta = E_{\text{TD-DFT}}(\text{ext}) - E_{\text{TD-DFT}}(\text{min})
\label{Eq.ext-min}
\end{equation}
was found to give reasonable results when combined with FS RCCSD energies (this model is denoted as ``CCSD+$\Delta$'' in Table~\ref{tab:exc-fscc}). The largest and mean absolute deviations for $5d$-levels are reduced to $-$0.31~eV and 0.18~eV, respectively (the PBE0 functional was employed). This level of accuracy can already be considered as quite acceptable, taking into account the overall difficulty of such a modeling and the problem under consideration as well as the wide manifold of possible sources of errors (basis set incompleteness, triple excitations, deficiencies of the CTEP approach itself). To ensure the stability of $\Delta$ corrections with respect to the particular choice of approximate exchange-correlation functional, the additional calculations with the PBE96 and HSE06 functionals were also performed. PBE96 results in a completely wrong picture of electronic states, while HSE06 gives nearly the same correction values as PBE0 (deviating by $<$0.03 eV). It should be pointed out that such a correction is justified only if one can unambiguously match the levels obtained from FS RCCSD and scalar-relativistic TD-DFT calculations. This is not so for the $4f$-levels of Ce$^{3+}$ and the corresponding cells in Table~\ref{tab:exc-fscc} are left empty. TD-DFT allows one to roughly estimate uncertainties for these levels arising mainly from the crystal field splitting, which were found to be $<$0.16 eV (MAD 0.08~eV). The spin-orbit splitting is not influenced by environment such dramatically for the deeply localized $f$-shells in the atomic core of Ce, since their spin-orbit interaction is also localized deeply in the atomic core. Our estimates for the uncertainties of the spin-orbit splitting of Ce $4f$-levels in YPO$_4$ is $\pm0.02$~eV. It should be pointed out that this error was not further reduced by increasing the fraction of the exact exchange in PBE0 to 50\%; thus, such large uncertainties for the $4f\to5d$ transition energies can be rather explained by overall deficiency of the DFT/PBE0 method.

\begin{table}
\caption{Energy levels (eV) of the Th$^{3+}$ impurity ion in xenotime matrix and corresponding excited state lifetimes calculated at the FS RCCSD level of theory. Lifetimes for states (2)--(8) exceed 1~$\mu$s and are not given here since large uncertainties for them are expected.}
\label{tab:exc-fscc-th}
\renewcommand{\arraystretch}{1.3}
\begin{tabular*}{\columnwidth}{l@{\extracolsep{\fill}}ccl}
\hline
\hline
No. & E, eV & $\tau$, ns & Composition \\
\hline
 1 & 0.00 & -- & $6d_{3/2}^{0.4}$ $6d_{5/2}^{0.4}$ \\
 2 & 0.53 & -- & $5f_{5/2}^{0.6}$ $5f_{7/2}^{0.1}$ $6d_{5/2}^{0.1}$ \\
 3 & 0.57 & -- & $5f_{5/2}^{0.4}$ $5f_{7/2}^{0.2}$ $6d_{3/2}^{0.1}$ $6d_{5/2}^{0.1}$ \\
 4 & 0.58 & -- & $5f_{5/2}^{0.4}$ $5f_{7/2}^{0.2}$ $6d_{3/2}^{0.1}$ $6d_{5/2}^{0.2}$ \\
 5 & 0.82 & -- & $5f_{5/2}^{0.1}$ $5f_{7/2}^{0.2}$ $6d_{3/2}^{0.3}$ $6d_{5/2}^{0.1}$ $7s_{1/2}^{0.1}$ \\
 6 & 0.95 & -- & $5f_{5/2}^{0.2}$ $5f_{7/2}^{0.5}$ $6d_{5/2}^{0.1}$ \\
 7 & 0.99 & -- & $5f_{5/2}^{0.3}$ $5f_{7/2}^{0.4}$ $6d_{5/2}^{0.1}$ $7s_{1/2}^{0.1}$ \\
 8 & 1.03 & -- & $5f_{5/2}^{0.1}$ $5f_{7/2}^{0.6}$ $6d_{5/2}^{0.1}$ \\
 9 & 2.34 & 434 & $5f_{5/2}^{0.1}$ $5f_{7/2}^{0.4}$ $6d_{3/2}^{0.1}$ $6d_{5/2}^{0.1}$ $7s_{1/2}^{0.1}$ \\
10 & 2.48 & 386 & $5f_{5/2}^{0.2}$ $5f_{7/2}^{0.3}$ $6d_{5/2}^{0.1}$ \\
11 & 2.56 & 272 & $5f_{5/2}^{0.2}$ $5f_{7/2}^{0.4}$ $6d_{5/2}^{0.2}$ \\
12 & 3.63 & 43.1 & $5f_{5/2}^{0.1}$ $5f_{7/2}^{0.1}$ $6d_{3/2}^{0.1}$ $6d_{5/2}^{0.2}$ \\
13 & 4.04 & 75.5 & $5f_{5/2}^{0.1}$ $5f_{7/2}^{0.2}$ $6d_{3/2}^{0.1}$ $6d_{5/2}^{0.2}$ $7s_{1/2}^{0.1}$ \\
14 & 5.28 & 3.02 & $5f_{7/2}^{0.1}$ $7p_{1/2}^{0.3}$ $7p_{3/2}^{0.1}$ \\
15 & 5.51 & 2.56 & $5f_{7/2}^{0.1}$ $7p_{1/2}^{0.3}$ $7p_{3/2}^{0.1}$ \\
16 & 6.01 & 6.46 & $7p_{3/2}^{0.2}$ \\
\hline
\hline
\end{tabular*}
\end{table}

The situation is much more involved for the Th$^{3+}-$doped xenotime (Table~\ref{tab:exc-fscc-th}). Spin-orbit coupling and interactions with a crystal field are of the same order. 
%
%
Moreover, $5f$ and $6d$ states of Th$^{3+}$ are notably closer to each other in both energies and spatial distributions than the $4f$ and $5d$ states of Ce$^{3+}$. The interference of these factors results in a strongly entangled picture of both energy levels (Fig.~\ref{fig:df_levels}) and wavefunction compositions. In contrast to the case of Ce$^{3+}$-doped crystal, the ground state is of pure $6d$ character on Th$^{3+}$. Then the bunch of seven electronic states with pronounced $5f$ contributions comes. According to the results of projection population analysis, the first three states of this group can be associated with the atomic $5f_{5/2}$ state, while the other four roughly correlates with the $5f_{7/2}$ 
component. This conclusion is also supported by the energy gap between barycenters of these two multiplets (0.39~eV), which is comparable with the spin-orbit splitting of the $5f$ levels in Th$^{3+}$ (0.54~eV, see Table~\ref{tab:th-basis}). All these levels have significant population of the $6d$ one-electron states. For the following five levels (No. 9 -- 13) such a clear interpretation is not possible, while the last two of them (No. 12 and 13) have up to 30\% of density which can be attributed to the $6d$ states. Last but not least, the uppermost levels considered in the present paper (No. 14 -- 16) can be interpreted as the $7p$ levels of Th$^{3+}$. The levels at 5.28 and 5.51 eV are significantly delocalized (only 50\% of the unpaired electron populates the Th ion), but still can be considered as meaningful. For the last level (at 6.01 eV) the small occupation number of the $7p$-spinors (0.2) clearly indicates that this excitation is delocalized and thus cannot be well treated within the minimal cluster model. Note that there is no level which can be associated with the $7s$ state of an isolated ion; its population is ``blurred'' between several other electronic states. The decrease in the gross population at the Th center allows us to argue that minimal cluster model is not applicable to higher-energy excitations that can be rather explained by the delocalized nature of $6d$ and even $5f$ states in Th.

Due to the complicated structure of electronic states (except for the first one), the scalar-relativistic TD-DFT corrections to energy levels cannot be firmly evaluated and applied for the Th$^{3+}$ doped xenotime cluster model. However, the upper bounds for corresponding uncertainties can still be set. For the states with dominating $5f$ population this uncertainty was found to be $<$0.15~eV (MAD 0.07~eV), for the states with the $6d$ character $<$0.28~eV (MAD 0.22~eV) (estimates at the PBE0 level). These values are actually the same as for the Ce$^{3+}$ containing cluster. As it can be seen from Table~\ref{tab:th-basis}, the error arising from basis set incompleteness is expected to be $<$0.07~eV. Thus the upper bound on the overall uncertainty of the energies given in Table~\ref{tab:exc-fscc-th} can be roughly estimated as not exceeding 0.35~eV.

We finally turn to the calculation of excited state lifetimes. Experimental data on lifetimes exist only for the lowest $5d$ level of YPO$_4$:Ce$^{3+}$, $\tau_{exptl}$ = 12.5 ns~\cite{Lai:07} or 23 ns~\cite{Laroche:01}. Relativistic coupled cluster calculation gives the estimate $\tau_{theor} = 23.7$~ns (see Table~\ref{tab:exc-fscc}; the experimental values of transition energies were used to calculate partial decay rates). For the solid state physics domain this result should be regarded as an excellent agreement between theory and experiment. Since non-radiative decay channels always exist in crystals, the first experimental value (12.5~ns) seems to be more reasonable and accurate. The lifetimes of other excited states of the Ce-doped cluster as well as the Th-doped one were also calculated (see Table~\ref{tab:exc-fscc-th}), but cannot be verified by the direct comparison with experiment. Based on uncertainty estimates for the calculated energy levels in the Th-doped xenotime ($<$0.35~eV), one can expect that uncertainty for the corresponding radiative lifetimes presented in Table~\ref{tab:exc-fscc-th} would be about 30\% 
($\tau \sim 1/\Delta E^3 |d|^2$, and typical uncertainties in transition dipole moments $d$ calculated by the finite-order method~\cite{Zaitsevskii:ThO:23} are of order of several percents). Note that the technique employed in the present study allows one to obtain transition moments for all pairs of electronic states in a single calculation~\cite{Zaitsevskii:ThO:23}. To the authors' best knowledge the coupled cluster theory has never been used previously to calculate lifetimes of excited states in solids.

\section{Conclusion}

A method of {\it ab initio} calculations of local properties in ionic crystals (equilibrium geometries of impurity sites, vertical excitation energies of impurity ions and excited-state radiative lifetimes) via constructing minimal cluster models with broken covalent bonds is proposed. The series of cluster models of pure xenotime and that doped with the Ce$^{3+}$ and Th$^{3+}$ impurity ions was constructed within the compound-tunable embedding potential (CTEP) approach. Relativistic coupled cluster method with the TD-DFT based corrections for the cluster model size
reduction
was shown to be able to reproduce energy levels of the Ce$^{3+}$ impurity ion in xenotime with the error not exceeding 0.31~eV (mean absolute deviation 0.18~eV) for $4f{\to}5d$ type transitions. The corresponding uncertainty for the thorium doped crystal was estimated to be less than 0.35~eV taking into account that $5f$ and $6d$ states are strongly mixed in general. The bulk of these uncertainties is due to the minimal cluster model and basis set incompleteness.
It should be pointed out that the estimated errors for the CTEP/RCC method developed in the present work are 3-4 times less than errors obtained for these systems within the TD-DFT approach ($> 1$~eV for several tested GGA, hybrid-GGA and meta-GGA functionals).

The technology of simulating impurity centers in crystals presented in the paper can be readily applied to systems with single $d$- or $f$-element impurity ion to obtain the information on its properties with relatively high and controlled accuracy unattainable for the TD-DFT method. The list of properties accessible within the CTEP+RCC framework includes not only electronic excitation energies, but also transition moments, properties in excited states, electronic factors for the $\mathcal{P}$- and $\mathcal{T},\mathcal{P}$-odd interaction operators, chemical shifts of lines of X-ray emission spectra.

It should be emphasized that the FS RCC method used here allows one to treat electronic states of impurity ions with up to three unpaired electrons (and/or holes)~\cite{Eliav:Review:22}, and is a method of choice for systems possessing strong multireference character like some lanthanide 
(Pr$^{3+}$, 
Nd$^{3+}$) and actinide (U$^{4+}$, U$^{3+}$, PuO$_2^{2+}$, etc) ions. Note that for most such objects the DFT/TD-DFT methodology is not applicable at all, 
since even ground electronic states of such systems possess strong non-dynamic correlations; furthermore, some electronic excitations can be of a two-electron nature.
Similar problems arise for relativistic versions of both the algebraic-diagrammatic construction~\cite{Pernpointner:18} and the conventional (particle-hole) equation of motion CC. However, for FS RCC the scope of applicability is also limited, and the development of fully relativistic versions of other powerful multireference methodologies like those presented in~\cite{Samanta:14,Aoto:16,Giner:16} is still highly desirable to treat the most challenging cases, like impurities of lanthanide and actinide ions with $f^4$~--~$f^{11}$ configurations.

Further progress in the CTEP+RCC framework can also include studies of processes involving two or more impurity atoms, e.~g., charge and photon transfer, fluorescence quenching, etc. Multi-center cluster models needed for such simulations would be at least two time larger in terms of the number of correlated electrons than the models considered in the present study.
Thus both further software improvements, including more sophisticated parallelization algorithms, and theoretical developments are needed.

\section{Acknowledgements}

Authors are grateful to Roman V.\ Bogdanov, Igor V.\ Abarenkov and Leonid V.\ Skripnikov for useful discussions.
Electronic structure calculations have been carried out using computing resources of the federal collective usage center Complex for Simulation and Data Processing for Mega-science Facilities at National Research Centre ``Kurchatov Institute'', http://ckp.nrcki.ru/.

The work
was supported by the Russian Science Foundation under grant no.~20-13-00225, https://rscf.ru/project/23-13-45028/.

\bibliography{JournAbbr,bibfile}

\begin{thebibliography}{97}
\bibitem [{Ana()}]{AnalyticDM:22}%
  \BibitemOpen
  \href@noop {} {}\bibinfo {note} {The new code for analytic density matrix
  calculations was developed and included into the EXP-T program package. The
  program is Kramers-unrestricted and allows one to generate analytic density
  matrices for single-reference open-shell CCSD and CCSD(T)
  models.}\BibitemShut {Stop}%
\end{thebibliography}

\end{document}